\documentclass[journal=nalefd,manuscript=article]{achemso}
\usepackage[utf8]{inputenc}
\usepackage{graphicx}
\usepackage{epstopdf}

\usepackage{color}
\usepackage{dcolumn}
\usepackage{bm}
\usepackage{amsmath}
\usepackage{amssymb}
\usepackage{hyperref}

\author{L. Iglesias}
\affiliation{Centro Singular de Investigación en Química Biolóxica e Materiais Moleculares (CIQUS), Departamento de Química-Física, Universidade de Santiago de Compostela, 15782, Santiago de Compostela, Spain}
\author{A. Gómez}
\affiliation{Institut de Ciència de Materials de Barcelona (ICMAB-CSIC), Campus UAB,  Bellaterra, Catalonia, 08193, Spain}
\email{a.gomez@icmab.es}
\author{M. Gich}
\affiliation{Institut de Ciència de Materials de Barcelona (ICMAB-CSIC), Campus UAB,  Bellaterra, Catalonia, 08193, Spain}
\author{F. Rivadulla}
\affiliation{Centro Singular de Investigación en Química Biolóxica e Materiais Moleculares (CIQUS), Departamento de Química-Física, Universidade de Santiago de Compostela, 15782, Santiago de Compostela, Spain}
\email{f.rivadulla@usc.es}

\title[Sample title]{Tuning oxygen vacancy diffusion through strain in SrTiO$_3$ thin films}


\begin{document}

\begin{abstract}

Understanding the diffusion of oxygen vacancies in oxides under different external stimuli is crucial for the design of ion-based electronic devices, improve catalytic performance, etc. In this manuscript, using an external electric field produced by an AFM tip, we obtain the room-temperature diffusion coefficient of oxygen-vacancies in thin-films of SrTiO$_3$ under compressive/tensile epitaxial strain. Tensile strain produces a substantial increase of the diffusion coefficient, facilitating the mobility of vacancies through the film. Additionally, the effect of tip bias, pulse time, and temperature on the local concentration of vacancies is investigated. These are important parameters of control in the production and stabilization of non-volatile states in ion-based memresistive devices. Our findings show the key role played by strain for the control of oxygen vacancy migration in thin-film oxides.

\end{abstract}

\maketitle

\section{Introduction}

Strong electronic correlations in narrow $d$-bands, along with a large coupling between charge, spin and orbital degrees of freedom, are responsible for the variety of electronic and magnetic behaviors characteristic of transition metal (TM) oxides (\textit{i.e.} metal-insulator transitions, superconductivity, ferroelectricity, multiferroicity, etc.). \cite{Goodenough2014,Khomskii2014, Goodenough1963} Of particular interest within them are the oxoperovskites, given the enormous flexibility of this structure to accept chemical substitution at the cationic sites. However, the possibility to control the occupancy of the anion site in an oxide is equally interesting, although much less explored. In this regard, oxygen vacancies behave in many 3$d$-oxoperovskites as mobile, $e$-donor defects, able to induce profound changes in the structural, electronic and magnetic properties of the material \cite{Schooley1964,Hu2016,Zhu2016}. 

Therefore, the possibility to control at will not only the creation of these vacancies, but also their distribution and movement inside the material using different external stimuli, is key for the design of ionic-based devices.\cite{Kalinin2013}

Strontium titanate, SrTiO$_3$ (STO) is a paradigmatic example in this regard. This oxoperovskite is a quantum paraelectric insulator\cite{Rowley2014}, whose great chemical stability, diamagnetism, and high dielectric constant makes it the preferred template for the growth of thin-films of other TM oxoperovskites. Incorporation of oxygen defects may be detrimental for some applications of this oxide; for instance, high-k requirements of dielectric insulating layers for carriers density modulation can be compromised.\cite{Fuchs1999,Yamada2005} However, they can be also used to add new functionalities to STO, like for instance gas sensing capabilities\cite{Hodak2010}, resistive switching\cite{Li2016,Pan2016,Janousch2007,Wu2014,Moon2017,Park},  ferroelectricity,\cite{Haeni2004,Jang2010} or low temperature superconductivity.\cite{Schooley1964}

In particular, resistive (switching) random-access memories (ReRAM) with controlled multi-stable states are attracting a considerable attention in recent years.\cite{Waser2009,Waser2007,Sawa2008} The distribution of V$_{\ddot{O}}$ in these devices can be modified by electrical or mechanical methods, resulting in reversible switching of their electrical resistance. Recently Kalinin et al.\cite{Das2017} discussed the electrical and mechanical manipulation of the local V$_{\ddot{O}}$ distribution in unstrained STO thin films using an AFM tip.  In a similar approach, Sharma et al.\cite{Sharma2015} were able to modulate the electrical conductivity of the 2D electron gas at the LaAlO$_3$/SrTiO$_3$ interface, opening a door towards mechanically operated transistors.

Here we report the effect of epitaxial strain in the diffusion coefficient, $D$, of V$_{\ddot{O}}$ at room temperature. We followed the time dependence of the local surface potential, after the distribution of V$_{\ddot{O}}$ is locally modified by an electric field produced by an AFM tip. Fitting of the experimental data to Fick's second law of diffusion demonstrates that tensile strain enhances $D$ considerably, while moderate compression does not show an appreciable effect with respect to unstrained films. We also studied the effect of the tip bias, pulse time, and temperature in the distribution of vacancies. These factors determine the control over  V$_{\ddot{O}}$ redistribution, the writing and retention times of the devices and constitute important factors for the applicability of ionic-based memresistive devices.


\section{Experimental Details}

SrTiO$_3$ thin-films (with $\approx2\%$ Nb $^{5+}$ to compensate the presence of unintentional Sr$^{2+}$ e-trap vacancies)\cite{Sarantopoulos,Iglesias2017} were grown by Pulsed Laser Deposition (PLD) (KrF, $\lambda$=248 nm and Nd:YAG $\lambda$=266 nm). The thin films were deposited on top of different substrates with pseudocubic lattice parameters ranging from 3.868$\textup{\AA}$ of (LaAlO$_{3}$)$_{0,3}$-(Sr$_{2}$AlTaO$_{6}$)$_{0,7}$ (LSAT) to 3.989$\textup{\AA}$  of KTaO$_3$ (KTO), in order to apply an epitaxial stress from -0.95$\%$ to +2.15$\%$. All samples used in this work were grown in the same batch, at 800\textordmasculine{}C and 100 mTorr of oxygen pressure,  with a thickness $\approx$17 nm.

The thin films were characterized combining Electrostatic Force Microscopy (EFM), and Kelvin Probe Force Microscopy (KPFM). The experiments were performed independently using two different AFM setups, a Park Systems NX10 and a Keysight 5500 AFM under both, ambient and low humidity conditions (6-7$\%$ RH), in order to suppress possible effects derived from charge injection\cite{Villeneuve-Faure2016}, formation of protons/hydroxyl radicals\cite{Rowicka2008},or electrochemical processes\cite{Collins2016}. A  Pt/Ir-coated metallic tip with force constant of 3 N/m was used for electrical scans in KPFM and EFM modes. The first resonance frequency of the lever was used to map the topography while the KPFM feedback was fed with the Xcomponent signal, and an excitation of 20kHz and 2VAC was applied to the tip. While KPFM is performed in single scan mode using tapping mode to acquire the topography, we performed the recording in contact mode to ensure a continuous electrical contact between the tip and the sample. We employed single-pass out of resonance KPFM, which fastened our data acquisition while avoid crosstalk between first and second resonance mode. The feedback of the KPFM is proportional to the Contact Potential Difference, VCPD, between the conducting cantilever and the sample surface. Such VCPD can be used to map the surface potential (SP) of the samples, which is known to be related to differences in charge density distribution\cite{Melitz2011}.


\section{Results and Discussion}

\begin{figure}[H]
\begin{center}
\includegraphics[width=350pt]{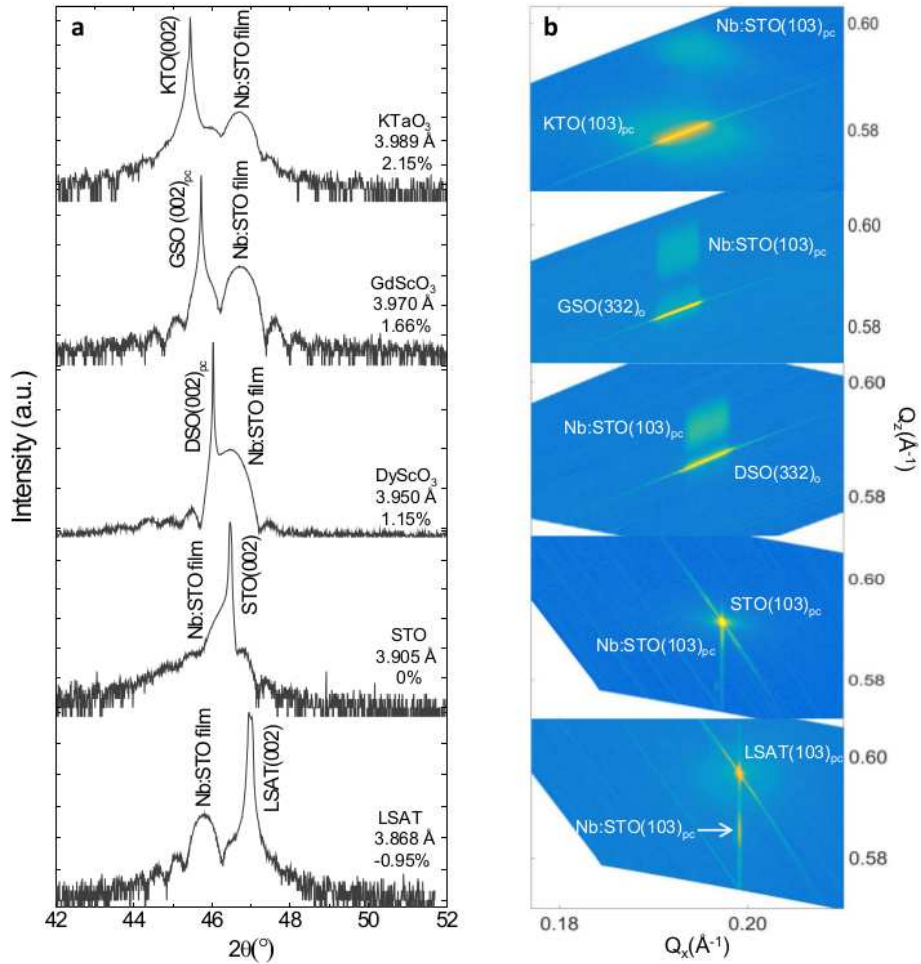}
\caption{a) XRD patterns around the (002) pseudo-cubic reflection. The lattice parameter of the different substrates, as well as the theoretical lattice mismatch with the film, is shown in each panel. b) High-resolution RSM around (103)$_{pc}$ asymmetric reflection in the case of LSAT, STO and KTO substrates, and (332)$_{o}$ reflection for DSO and GSO substrates.}
\label{fig:1}
\end{center}
\end{figure}

The structural properties of the films were studied by high-resolution X-Ray Diffraction (XRD), and are summarized in Figure \ref{fig:1}. The X-Ray Reciprocal Space Maps (RSM) of the films confirm that they are fully strained to the in-plane lattice parameter of the substrates. The Laue oscillations around the (002) diffraction peak also prove that the films maintain a very good structural quality, irrespectively from the sign and magnitude of the lattice mismatch with the substrate. 

The epitaxial compressive (tensile) strain induced by the substrate changes the in-plane lattice parameter of the films (from -0.95$\%$ on LSAT, to +2.15$\%$ on KTO), and causes an expansion (contraction) along the out-of-plane lattice parameter. This deformation of the unit cell can be calculated assuming an elastic deformation given by the Poisson's ratio of $\nu$=0.2332\cite{Ledbetter1990} (see Fig. S1 in the supporting information). The evolution of unit cell volume follows the prediction of $\nu$, therefore confirming that there are no important deviations in cation stoichiometry, which could compromise the analysis of the results.

\begin{figure}[H]
	\begin{center}
		\includegraphics[width=300pt]{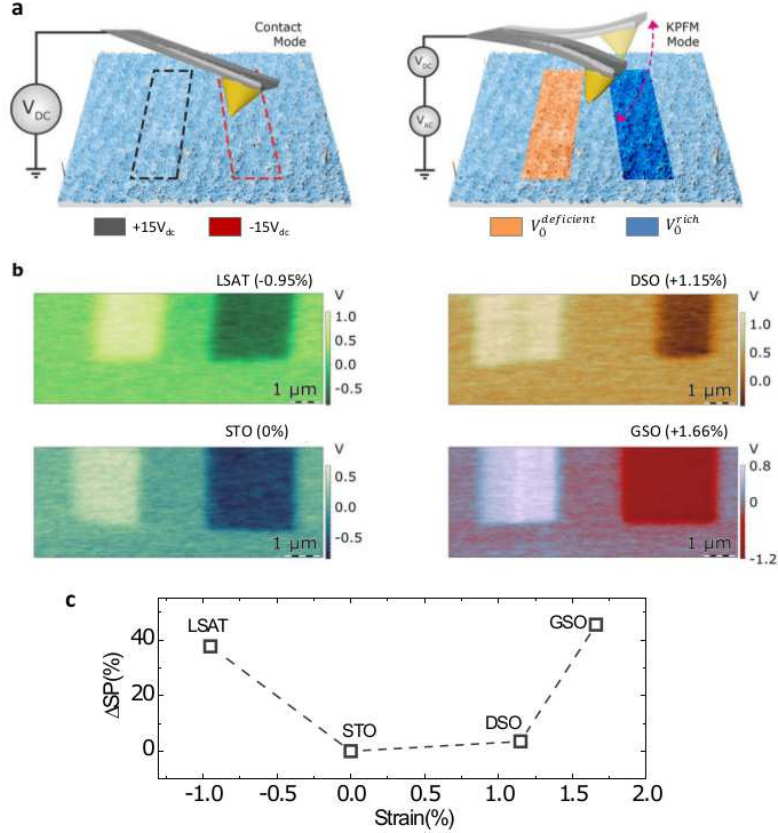}
		\caption{a) Sketch of the KPFM method to record the sample surface with positive and negative tip bias. Positive bias produces a deficient V$_{\ddot{O}}$ concentration, while a negative voltage leads to an enrichment of the local concentration of V$_{\ddot{O}}$. b) Characterization of the surface potential for the thin films grown on different substrates. c) Evolution of $\Delta$SP as a function of epitaxial strain. $\Delta$SP is defined as the difference between the minimum/maximum in the surface potential signal of the V$_{\ddot{O}}$ enriched/depleted regions.}
		\label{fig:2}
	\end{center}
\end{figure}

 In order to quantify the effect of epitaxial strain on the drift mobility of V$_{\ddot{O}}$, a region on the surface of each sample was first poled with negative/positive AFM tip bias in contact mode, and then the exact same area was mapped by KPFM scanning at zero voltage. An schematic of this process is depicted in Figure \ref{fig:2}a. The highly localized electric field ($\sim$6MV/m) generated by the AFM bias, ineherent to the nanometer scale radius of the tip, is sufficient to modify the initial homogeneous distribution of positively charged V$_{\ddot{O}}$ across the film. Depending on the electric field direction, the accumulation or depletion of V$_{\ddot{O}}$ produces a localized increment or decrease on the SP of the sample \cite{Girard2001}.

A negative voltage causes an accumulation of vacancies, which contribute with free electrons to the sample surface, producing a local decrease of the work function and therefore of the surface potential. On the contrary, a positive tip bias drives V$_{\ddot{O}}$ away from the sample surface increasing SP.

Significant differences in the background surface potential of the unpolarized regions are evident, depending on the magnitude of the strain. This is related to a slight different initial concentration of V$_{\ddot{O}}$ in the as-grown films, and therefore, to the different initial conductivity of the samples\cite{Kubicek2013}. To minimize artifacts and offsets from the measurement conditions, all measurements were performed with the exact same probe, AC bias magnitude, frequency, tip-sample distance, and scanning/recording rate for both KPFM and topography acquisition.

Following this procedure, we are able to compare how the relative change of the surface potential between V$_{\ddot{O}}$ enriched/depleted regions, $\Delta$SP, changes with strain (Figure \ref{fig:2} b). $\Delta$SP increases up to $\approx$40$\%$ for the most stressed films, either compressive (on LSAT) or tensile (on GSO) (Figure \ref{fig:2} c)), with respect to the unstrained film (on STO). Therefore, epitaxial strain influences very much the drift mobility and the subsequent accumulation/depletion of V$_{\ddot{O}}$ in STO.

\begin{figure}[H]
\begin{center}
\includegraphics[width=300pt]{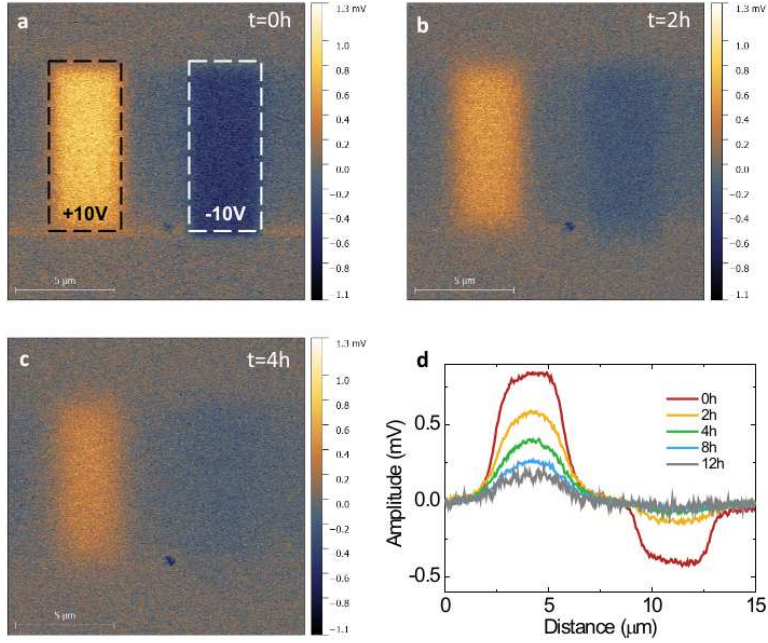}
\caption{Evolution of the EFM amplitude over time at zero electric field. a) Immediately after removing the AFM tip bias, b) after 2 hours, and c) after 4 hours. d) Profile of the EFM amplitude recorded over 12 hours. The results correspond to a film of STO on GSO. The effect of tip bias in the surface potential is completely reversible, independently of the sign and magnitude of strain (Figure S2 of Supplementary Information).}
\label{fig:3}
\end{center}
\end{figure}

 In order to determine the diffusion coefficient of the oxygen vacancies, first an area of the sample is poled with an electric field of $\pm$10 V. Then, the electric field is set at 0 V, and the time dependence of the surface potential (by monitoring the EFM and KPFM signal) is followed for one day, every 30 min. The vacancies diffuse freely after the field is removed, blurring the effect of the charge accumulation/depletion on the surface potential  (Figure \ref{fig:3} a),b) and c)).

The profiles at different times were extracted from the EFM images, see Figure \ref{fig:3} d). These changes reflect the diffusion of V$_{\ddot{O}}$, which are expected to follow the behavior predicted by second Fick's law to attain an equilibrium distribution:

\begin{equation}\label{eq1}
\frac{\partial {[V_{\ddot{O}}]}}{\partial t}=D\frac{\partial^2 {[V_{\ddot{O}}]}}{\partial^2 z^2}
\end{equation}

This equation can be solved under certain boundary conditions. If a predominant diffusion along the out-of-plane direction (z) is assumed, the diffusion equation can be solved in the limit of t${\rightarrow\infty}$(see the Supplementary Information for a complete deduction of this expression)\cite{Crank1975,Merkle2008}:

\begin{equation}\label{eq2}
\Delta V_{\ddot{O}}(t)=\frac{8}{\pi^2}e^\frac{-\pi^{2}Dt}{l^2}
\end{equation}

\begin{figure}[H]
	\begin{center}
		\includegraphics[width=400pt]{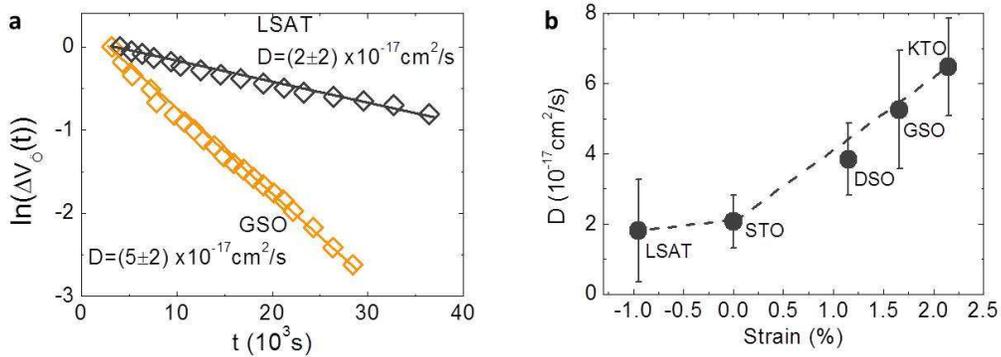}
		\caption{a) Time decay of the surface potential due to V$_{\ddot{O}}$ diffusion in STO grown on GSO and LSAT. Lines are fitting to equation (\ref{eq2}). b) Strain dependence of the diffusion coefficient of V$_{\ddot{O}}$. The dashed line is a guide to the eye.}
		\label{fig:4}
	\end{center}
\end{figure}

In this expression $D$ is the diffusion coefficient and \textit{l} is the film thickness (\textit{l}=17nm for all films used in this work). The fitting of the experimental data to equation (\ref{eq2}) is shown in Figure \ref{fig:4} a) for two representative cases. The value of $D$$\backsim$2x10$^{-17}$cm$^2$/s for the unstrained film is in very good agreement with the value reported in bulk STO at room temperature\cite{DeSouza2012,Das2017}. In contrast, 2\% of tensile strain increases $D$ by an impressive $\approx$350 \% with respect to the unstrained films (Figure \ref{fig:4} b). On the other hand, compressive strain does not show any appreciable effect on the difussion coefficient of oxygen vacancies in STO .

These results are fully consistent with DFT calculations which predicted a continuous increase of the V$_{\ddot{O}}$ diffusion with tensile strain,\cite{Rayson2013} while a compressive strain  larger than $\approx$-4$\%$ is necessary to decrease the energy barrier for diffusion \cite{Hamadany2013}. In fact, these calculations predicted a significant role of complex TiO$_6$ rotations in the anisotropic oxygen diffusion in STO. We recently observed that tensile strain induces an out-of-phase rotation of the TiO$_6$ octahedra along the c-axis, while compressive strain suppresses this rotation completely.\cite{Iglesias2017} Therefore, not only the magnitude of stress, but the changes in the internal rotations of TiO$_6$ octahedra could be playing a significant role in oxygen diffusion in STO thin-films.

The dramatic dependence of $D$ with strain has important implications for the design of ionic-conducting devices, like resistive switching memories: tensile strain would favour the writing process in such devices, but would also reduce the retention state due to the increase in the V$_{\ddot{O}}$ mobility. Therefore, ion-based devices of this type will benefit from compressive strain.

\begin{figure}[H]
	\begin{center}
		\includegraphics[width=250pt]{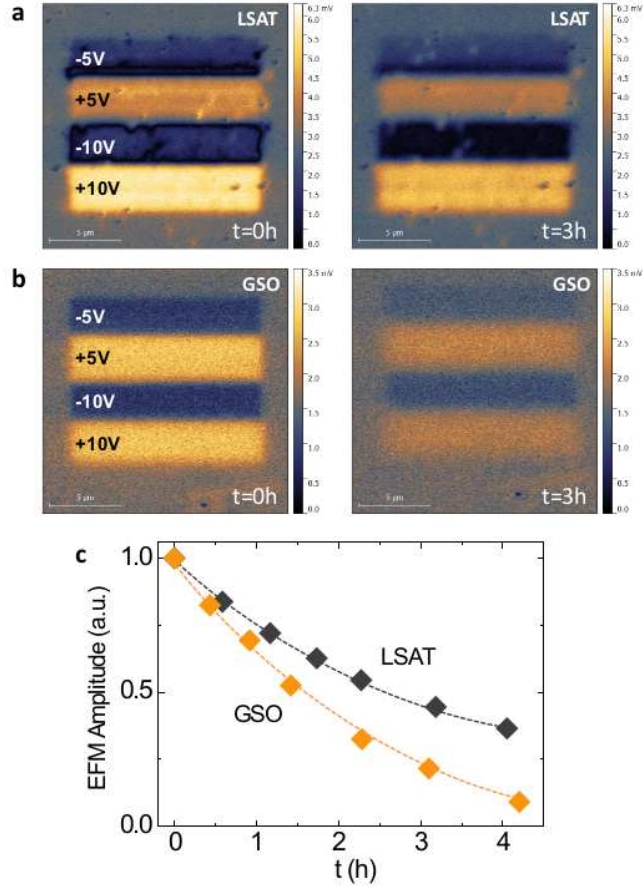}
		\caption{Record of different potential surface states with a variable voltage from -10V to 10V. a) EFM amplitude on LSAT immediately after the record and 3 hours later. b) Same record on the sample deposited on GSO. c) Normalized evolution of the EFM amplitude for both samples during 4.5 hours. After 3 hours the surface potential decays $\approx$50\% from the initial value in LSAT, and $\approx$75\% in GSO. Dashed lines are guides to the eye.}
		\label{fig:5}
	\end{center}
\end{figure}

To study the effect of diffusion and the possibility to define multiple stable neighboring states, we recorded side-by-side areas with different voltages, under compressive (LSAT, figure \ref{fig:5} a)) and tensile stress (GSO, figure \ref{fig:5} b)). The results demonstrate clear contrast boundaries between neighboring areas poled with $\pm$ voltages, with excellent retention between them. However, due to the effect of strain on $D$, the signal decays $\approx$50$\%$ after 3 hours on the sample grown on LSAT, and $\approx$75$\%$ in GSO (Figure \ref{fig:5} c)). This demonstrates again the advantage of compressive strain in the stabilization and retention of the multipotential surface states.

\begin{figure}[H]
	\begin{center}
		\includegraphics[width=380pt]{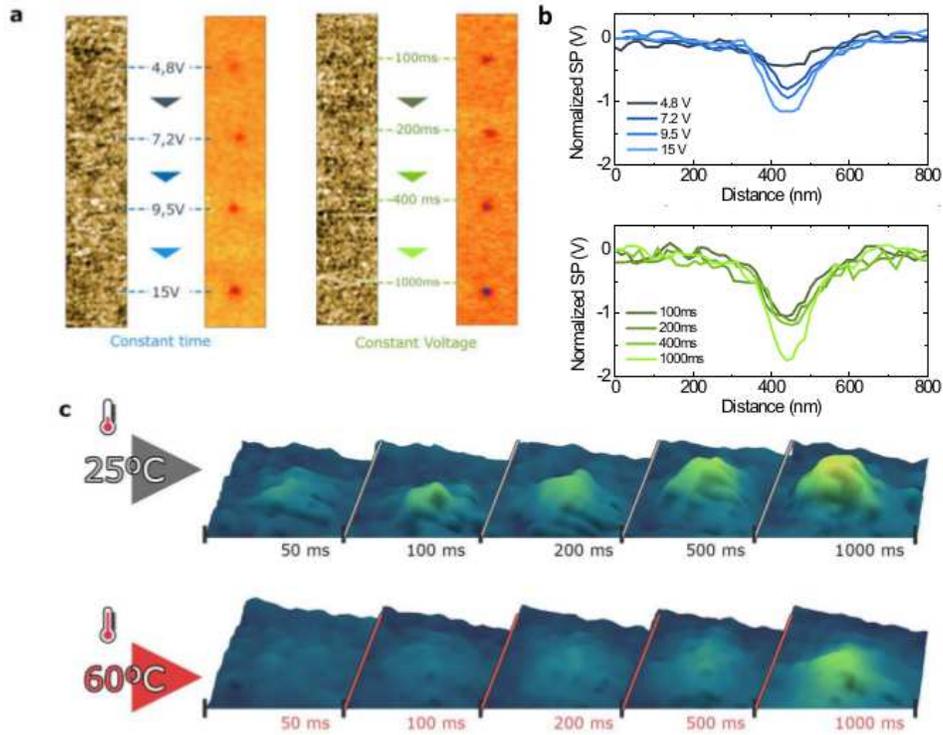}
		\caption{Influence of tip bias, pulse time and temperature on the V$_{\ddot{O}}$ distribution on GSO. a) Topography (blue) and surface potential (green) of four-dot array, which were performed by pausing the tip for 0.4s and varying the tip bias for 4.8V to 15V (left). Dots recorded while keeping the voltage constant at 15V and varying the pulse time for 0.1 to 1s (right). b) Resulting profiles of the normalized surface potential for the dots poling at constant time (up) and constant voltage (down), respectively. c) 3D-plot of the surface potential, related to the V$_{\ddot{O}}$ distribution, after different pulse times (+15 V) applied at 25\textordmasculine{}C and 60\textordmasculine{}C.}
		\label{fig:6}
	\end{center}
\end{figure}

Apart from low writing voltage/time, and high retention, competitive non-volatile ionic devices also require a high thermal stability. We performed successive poling experiments applying a constant pulse time with different tip bias, and viceversa, maintaining the tip in contact with the sample surface. The results for topography and KPFM surface potential at 25\textordmasculine{}C are shown in Figure \ref{fig:6} a). Note that the impact of  V$_{\ddot{O}}$ accumulation is negligible in the topography, irrespective of the tip bias or the pulse time. In Figure \ref{fig:6} b) the profile of the KPFM image evidences the existence of a threshold voltage ($\backsim$5-6 V) above which there is a notable increment of the local accumulation of V$_{\ddot{O}}$.  Similarly, a threshold pulse time of 500 ms increases significantly the local concentration of V$_{\ddot{O}}$. The existance of these threshold voltages/times indicates an activation energy for ionized vacancies to detach from lattice defects, most probably cationic vacancies, which can affect the drift mobility beyond strain.\cite{Szot2006}

Regarding the temperature dependence of the V$_{\ddot{O}}$ distribution, we compared write/read voltage measurements performed at room temperature (25\textordmasculine{}C) and at 60\textordmasculine{}C, using different pulse times with same tip bias of +15V (Figure \ref{fig:6} c)). Poling of the sample and subsequent KPFM scan are separated by 15 minutes. The required pulse time to achieve a measurable change of the surface potential increases substantially with temperature, reflecting the rapid increase of $D$(T). This result shows the low retention of the multipotential states, which can be erased at moderate temperatures.

In summary, we have shown the dramatic effect of epitaxial strain on the diffusion coefficient of V$_{\ddot{O}}$ in STO. Tensile strain enhances the mobility of V$_{\ddot{O}}$ through the films, being therefore detrimental to the stable definition of multi-stable states in memristors and similar devices. Additionally, we have proved that other parameters such as tip bias, pulse time and temperature can also control the V$_{\ddot{O}}$ distribution, offering an alternative tool to write/erase multi-stable states in ion-based devices. 


\begin{acknowledgement}
	This work was supported by the Ministerio de Economía y Competitividad of Spain under project numbers MAT2016-80762-R and SEV-2015-0496 (Severo Ochoa Program Grant), and Xunta de Galicia (Centro Singular de Investigación de Galicia accreditation 2016-2019) and the European Union (European Regional Development Fund – ERDF). L. I. also acknowledges the Ministerio de Economía y Competitividad of Spain for a FPI grant. 
\end{acknowledgement}

\providecommand{\latin}[1]{#1}
\makeatletter
\providecommand{\doi}
{\begingroup\let\do\@makeother\dospecials
	\catcode`\{=1 \catcode`\}=2 \doi@aux}
\providecommand{\doi@aux}[1]{\endgroup\texttt{#1}}
\makeatother
\providecommand*\mcitethebibliography{\thebibliography}
\csname @ifundefined\endcsname{endmcitethebibliography}
{\let\endmcitethebibliography\endthebibliography}{}

\end{document}